# Exploiting Electrical Transients to Reveal Charge Loss Mechanism of Junction Solar Cells

Yiming Li,[a,c†] Jiangjian Shi,[a†] Bingcheng Yu,[a,c] Biwen Duan,[a,c] Jionghua Wu,[a,c] Hongshi Li,[a,c] Dongmei Li,[a,c,d] Yanhong Luo,[a,c,d] Huijue Wu,[a] and Qingbo Meng[a,b,d*]

*a. Key Laboratory for Renewable Energy, Chinese Academy of Sciences; Beijing Key Laboratory for New Energy Materials and Devices; Institute of Physics, Chinese Academy of Sciences, Beijing 100190, P. R. China.*
*b. Center of Materials Science and Optoelectronics Engineering, University of Chinese Academy of Sciences, Beijing 100049, China.*
*c. School of Physics Science, University of Chinese Academy of Sciences, Beijing 100049, P. R. China.*
*d. Songshan Lake Materials Laboratory, Dongguan, Guangdong 523808, China.*

† Contributing equally to this work.

*E-mail: qbmeng@iphy.ac.cn

**Abstract:** Electrical transients enabled by optical excitation and electric detection provide a distinctive opportunity to study the charge transport, recombination and even the hysteresis of a solar cell in a much wider time window ranging from nanoseconds to seconds. However, controversies on how to exploit these investigations to unravel the charge loss mechanism of the cell have been ongoing. Herein, a new methodology of quantifying the charge loss within the bulk absorber or at the interfaces and the defect properties of junction solar cells has been proposed after the conventional tail-state framework is firstly demonstrated to be unreasonable. This methodology has been successfully applied in the study of commercialized silicon and emerging $Cu_2ZnSn(S, Se)_4$ and perovskite solar cells herein and should be universal to other photovoltaic device systems with similar structures. Overall, this work provides an alluring route for comprehensive investigation of dynamic physics processes and charge loss mechanism of junction solar cells and possesses potential applications for other optoelectronic devices.

# 1. Introduction

Functionalization of a semiconductor device, such as photovoltaic, integrated circuit and etc., is mainly enabled by optical-electrical or electrical-electrical inter-conversions and charge-carrier processes within the semiconductor material.[1-4] Recently, hybrid lead halide perovskite, which exhibits outstanding optical-electrical conversion abilities, has been identified to possess abundant charge-carrier characteristics, such as slow hot carrier cooling,[5-8] long polarization memory[9-11] and fast ion migrations[12-15]. These novel photophysical behaviors significantly enrich the semiconductor family and have spawned a new discipline, that is, perovskite optoelectronics. Probing and manipulating these processes provide an effective route for improving the material performance and extending the perovskite application.

For the perovskite solar cell, steady-state energy conversion involves a variety of processes including charge generation, charge transfer and transport within the bulk absorber and the charge transporting layers (CTLs), charge recombination at the bulk and interface, and even photoelectric hysteresis.[16] A series of optical and electrical transient methods has been exploited to probe these dynamics processes in a wide time ranging from sub-picosecond to even hours to shed light on the working mechanisms and charge loss mechanism of this emerging material and cell.[17-37] Compared to the advanced ultrafast optics measurements,[17-29] perturbation electrical transients (e.g. transient photocurrent (TPC) and photovoltage (TPV))[30-37] enabled by optical excitation and electric detection provide an opportunity to study charge transport, recombination and even the hysteresis in a much wider time window. This technique has been widely applied for silicon,[38-39] sensitized,[40-41] quantum dot,[42] organic[43] and recent the perovskite solar cells.[30-35]

A generic physics model and quantitative analysis method centered on the charge occupation of subgap tail states has been extended from sensitized to current perovskite solar cells.[31-32, 44-46] Within this model, the distribution of tail states has been interpreted as a common origin for the difference in the device performance and especially the open-circuit voltage, no matter the device structure and working mechanism. This raises several fundamental questions, that is, (1) Do all the cells follow the same electrical transient processes or charge loss mechanism? (2) Can the electrical transients be used to reflect the distribution of tail states or subgap states? (3) What does the electrical transient measurement really tell?

Controversies on the electrical transients for solar cells have always been ongoing. For instance, Frank Marlow et al. doubted whether the charge transport in dye-sensitized solar cells had been really understood, according to their voltage-dependent photocurrent measurement.[47] Recently, Kristofer Tvingstedt et al. also proposed a new understanding on the charge lifetime derived from the electrical transients of a variety of thin film photovoltaic devices.[30] For the perovskite solar cell, whether the measured tail states arise from the bulk perovskite or from the charge transporting layer is also under controversial.[48-50] Therefore, understanding the device physics truth from the electrical transients and exploiting these approaches to unravel the charge loss mechanism of solar cells, especially of the emerging perovskite solar cell, should be an important research topic for the photovoltaic field.

In this work, we propose a new measurement and analysis methodology to quantitatively extract charge dynamics properties and charge loss mechanism of photovoltaic devices from the electrical transients, such as charge extraction and collection quantum efficiency and the density of defects within the absorber. This methodology has been successfully applied to study conventional silicon and emerging kesterite and perovskite solar cells herein and is able to extend to other photovoltaic device systems with similar structures. Therefore, this work provides an alluring route for comprehensive investigation of dynamic physics processes and charge loss mechanism of junction solar cells and possesses potential applications for other optoelectronic devices.

For clarity, in this paper we will firstly focus on the differential capacitance of devices and discuss the validness of conventional tail state framework. Then, carrier dynamics and charge loss mechanism behind electrical transients will be further studied based on theoretical calculation. Finally, we will propose a new measurement and analysis methodology and self-consistent physics model to quantify charge loss and defect density by electrical transients, which is proved to be applicable for a series of junction solar cells. Notably, all the discussions in this work are limited within the perturbation electrical transient analysis, which may not suitable for other methods that significantly deviate from quasi-steady state.

## 2. Results and discussions

Figure 1(a) gives a schematic diagram of general charge-carrier processes within a perovskite solar cell where the perovskite absorber (PVSK) is sandwiched by electron (ETL) and hole transporting layers (HTL). Sn: $In_2O_3$ (ITO) and Au are used as the front and back electrodes to collect electrons and holes from the ETL and HTL, respectively. To produce electricity, the photon generated charge need to experience transportation in the PVSK, extraction at the ETL and HTL/perovskite interface and collection at the electrode interface. Due to the bulk and interface defects, charge loss is usually unavoidable. The cell performance is ultimately determined by the dynamic competition between charge transport-extraction-collection and charge loss due to recombination. As shown in Figure 1(b), these processes occur in a wide time scale ranging from nanosecond to millisecond, which is exactly within the detection window of electrical transients.[17-37] Therefore, electrical transient is expected to be a powerful approach to reveal the charge loss mechanisms of a solar cell.

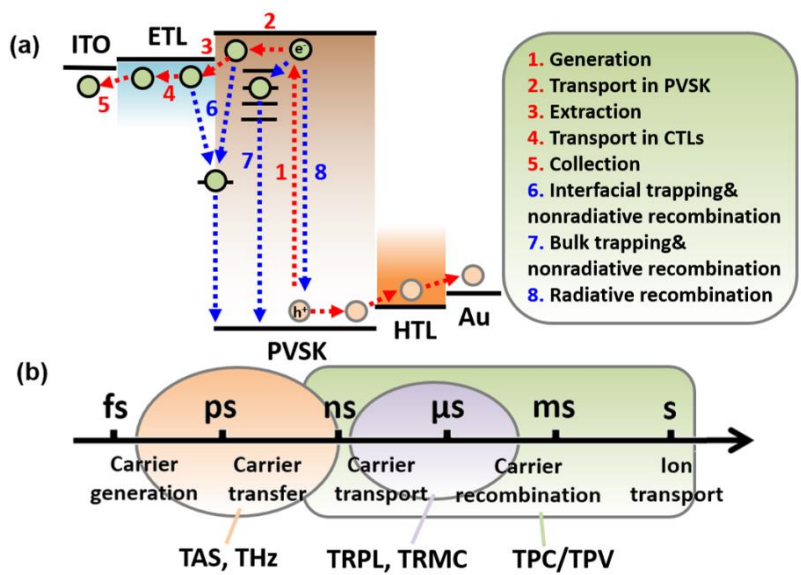

**Figure 1. General charge-carrier processes within a perovskite solar cell.** (a) Carrier dynamics processes and (b) the corresponding time scale of the cell. Different transient

technologies including transient absorption spectroscopy (TAS),[17-18] terahertz pump-probe technology (THz),[19-21] time-resolved photoluminescence (TRPL),[22-26] microwave photo-conduction (TRMC)[27-29] and TPC/TPV measurements[30-37] have been used to probe these carrier dynamic processes.

**Tail state framework – Differential capacitance**

Conventionally, charge loss mechanism of a solar cell derived from the TPC/TPV analysis is explored within the framework of tail state distribution.[31-32] For this framework, the variation in the device performance and especially in the open-circuit voltage ($V_{OC}$) is mainly attributed to the tail (or subgap) state distribution, which is usually reflected by the stored charge ($Q(V_{OC})$) at the open-circuit condition. The $Q(V_{OC})$ is obtained by integrating the differential capacitance ($C_{DC}$) of the cell and the $C_{DC}$ is determined by the TPC/TPV method, that is,

$$C_{DC}(V_{OC}) = \Delta Q(V_{OC})/\Delta V(V_{OC}) \approx \Delta Q(I_{OC})/\Delta V(V_{OC}), \qquad (1)$$

where $\Delta V$ is the TPV peak at different $V_{OC}$, $\Delta Q$ is the perturbation charge integrated from the TPC decay at different light illumination intensity ($I_{OC}$) corresponding to the $V_{OC}$. Generally, $\Delta V$ is measured in the open-circuit (OC) condition while $\Delta Q$ is measured in the short-circuit (SC) condition. An assumption is made to make equation (1) applicable, that is, the difference of $\Delta Q(I_{OC})$ to the $\Delta Q(V_{OC})$ can be ignored. Based on this assumption, this framework was widely used for the sensitized and organic solar cells and has been extended to the perovskite solar cells.[31-32, 44-46] In the following section, we will firstly check the validness of this assumption based on the $C_{DC}$ measurement of perovskite solar cells.

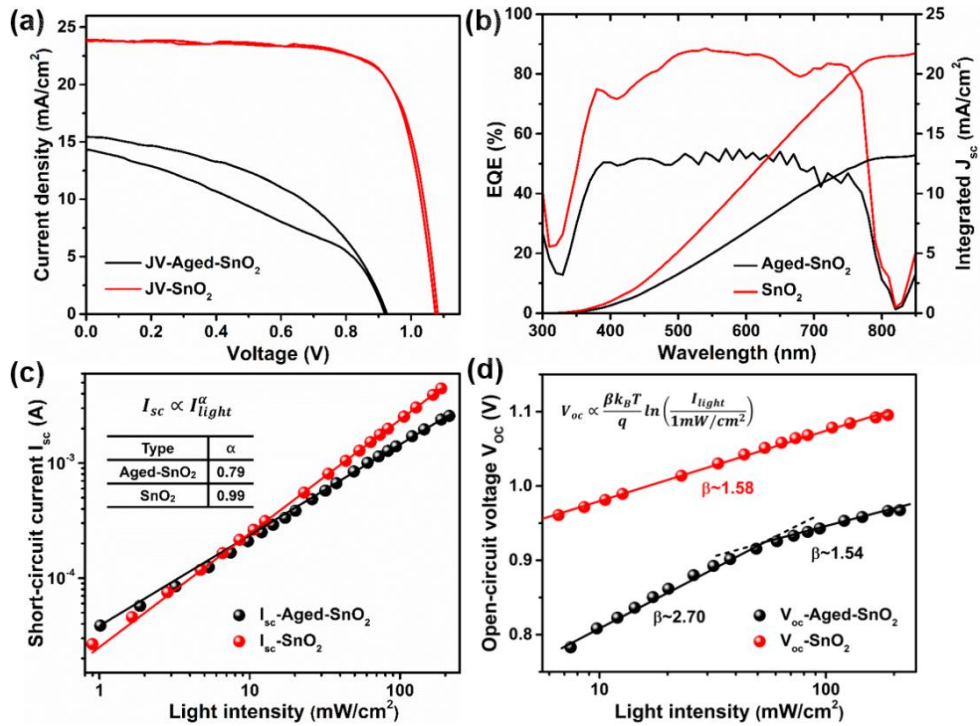

**Figure 2. Current-voltage (*I-V*) and charge recombination properties of the studied cells.** (a) Representative *I-V* curves of the SnO$_2$ based perovskite solar cell without or with experiencing time aging. (b) EQE spectra of the cells. Light intensity-dependent (c) short-circuit current ($I_{sc}$) and (d) open-circuit voltage ($V_{OC}$).

A perovskite solar cell with a configuration of ITO/SnO$_2$/PCBA/perovskite/Spiro-OMeTAD/Au was used as the standard sample, which gives power conversion efficiency (PCE) of 19.2% with negligible hysteresis behaviour. The PCBA was used here as an interface passivation layer, which can help significantly suppress the interface defect.[51-52] With this passivation, no signature of severe interface recombination (i.e. fast early-stage TPV decay) has been observed in our experiment here. Thus, possible charge loss and bending of quasi-Fermi energy level splitting (QFLS) at interfaces will not be considered in this work.[53] Another sample with a significantly degraded PCE of ~7% was obtained by aging the standard cell for several months in ambient conditions. Notably, phase segregation of the perovskite absorber that could obviously degrade the cell[54] did not appear in our aged sample. Thus, we infer that the efficiency degradation here mainly arise from the variations of the defect and charge recombination properties. Figure 2(b-d) gives bias light intensity dependent SC current and $V_{OC}$ characterizations of these two cells to further evaluate the degradation mechanism. The $α$ values of the fitted line for Aged-SnO$_2$ and SnO$_2$ devices presented in Figure 2(c) are 0.79 and 0.99, respectively, indicating that the Aged-SnO$_2$ device exhibits a higher recombination loss under short-circuit condition.[55] In Figure 2(d), the obtained $β$ of the SnO$_2$ device is around 1.6, which is consistent with previous results[56]. For the Aged-SnO$_2$ device, the obtained $β$ is higher than 2.0 when the light intensity is lower than 60 mW cm$^{-2}$, suggesting a higher defect recombination loss. At higher light intensity, the $β$ decreases to 1.5. This implies that the defects have been filled by the photo-induced charge with charge recombination being suppressed more or less.

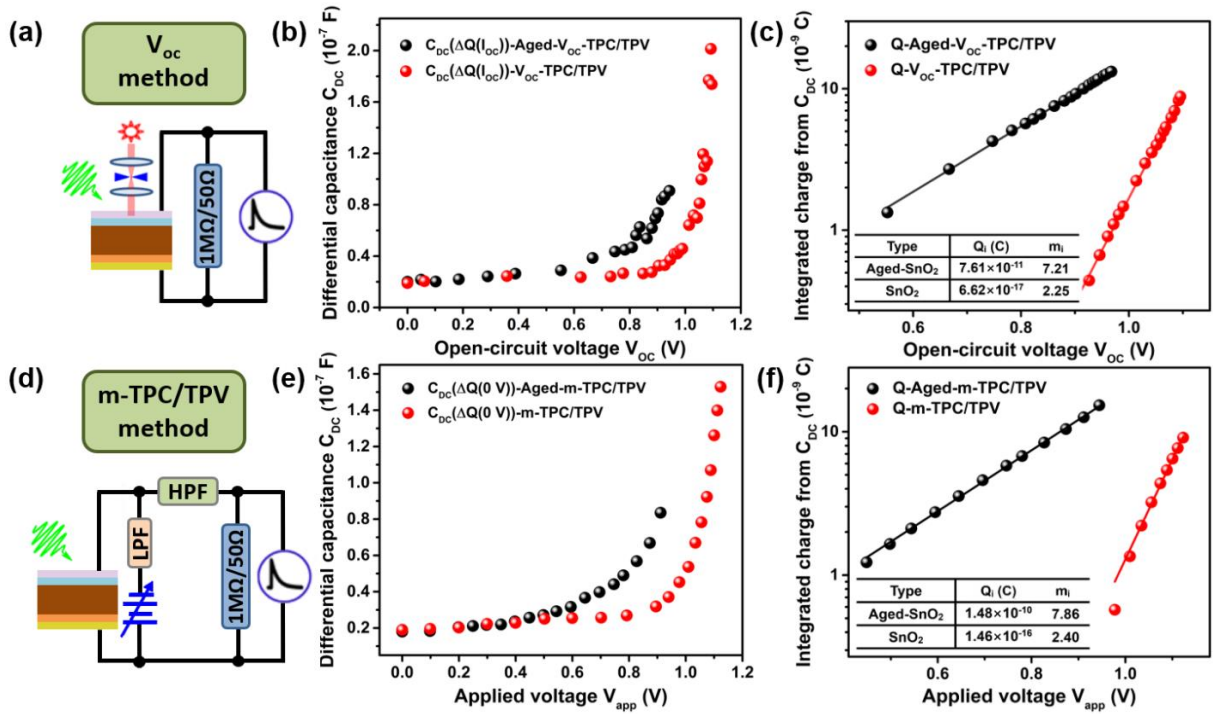

**Figure 3. Evaluating the tail state of the cell within the conventional framework.** (a) Schematic diagram of the $V_{OC}$-TPC/TPV measurement. An LED with adjustable light intensity is used to produce the steady state $V_{OC}$ voltage bias. (b) Differential capacitance $C_{DC}$ as a function of $V_{OC}$ for the Aged-SnO$_2$ and SnO$_2$ devices. (c) Total integrated charge density Q as a function of $V_{OC}$. (d) Schematic diagram of the designed M-TPC/TPV system. A steady-state voltage source is used to give bias voltage to the cell and a high-pass filter (HPF) together with a low-pass filter (LPF) is applied to separate the transient electrical signal from the voltage source to avoid shunting. (e) $C_{DC}$ as a function of applied voltage $V_{app}$ for the cells. (f) $Q$ in two devices as a function of $V_{app}$. All the theoretical fittings were made within the conventional tail-state framework.

We firstly use the conventional tail state scenario and equation (1) to evaluate the $C_{DC}$, and thus to compute the stored charge ($Q$ ($V_{OC}$)) within the cell (especially within the tail states) under varied $V_{OC}$ with an integration of the $C_{DC}$, that is,

$$Q(V_{OC}) = \int_0^{V_{OC}} (C_{DC} - C_{electrode}) dV \qquad (2)$$

where $C_{electrode}$ is the electrode or geometry capacitance determined from the flat baseline of $C_{DC}$ at low $V_{OC}$. This $V_{OC}$-TPC/TPV measurement scheme is shown by Figure 3(a), where the TPC is measured with a 50 Ω sampling resistor and TPV is measured with a 1 MΩ sampling resistor. Figure 3(b-c) gives the $C_{DC}$ and integrated $Q$ results, which exactly agree well with the previous scenario with an exponential distribution feature.[31] The tail state properties of the perovskite was believed to be able to be parameterized by fitting the distribution of $Q$ as

$$Q(V_{OC}) = Q_i \exp\left(\frac{qV_{OC}}{m_i K_B T}\right), \qquad (3)$$

where $q$ is the elementary charge, $K_B$ is the Boltzmann constant, $T$ is the absolute temperature, $Q_i$ and $m_i$ are the fitting factors which are usually used to reflect the effective density of states of the tail state. Clearly, degraded sample gives a much larger $Q_i$, which was believed to indicate a much higher tail state density according to the conventional scenario. This interpretation seems reasonable because higher tail state density can indeed lead to severe charge recombination and lower cell efficiency.

In fact, besides the bias light induced $V_{OC}$, we can also adjust the electric properties of the cell by directly applying a paralleled-connected external voltage source. To avoid the influence of this voltage source on the TPC/TPV detecting, a low-pass filter (LPF) is used to separate the high-frequency transient signal, as shown in Figure 3(d), which leads to the realization of instrument design of modulated TPC/TPV (m-TPC/m-TPV), as we once reported.[33] By using

$C_{DC}(V_{app}) \approx \Delta Q (0\ V)/\Delta V (V_{app})$, where $V_{app}$ is the applied bias voltage, a similar result to that of tail state scenario can be obtained, as shown in Figure 3(e-f). This indicates that $V_{OC}$ plays a similar role to the $V_{app}$ in influencing the cell electronic properties and more importantly that our previous instrument design itself cannot bring new insight into the characterization of electrical transient and charge loss mechanism of solar cells if still following the conventional analysis scenario.

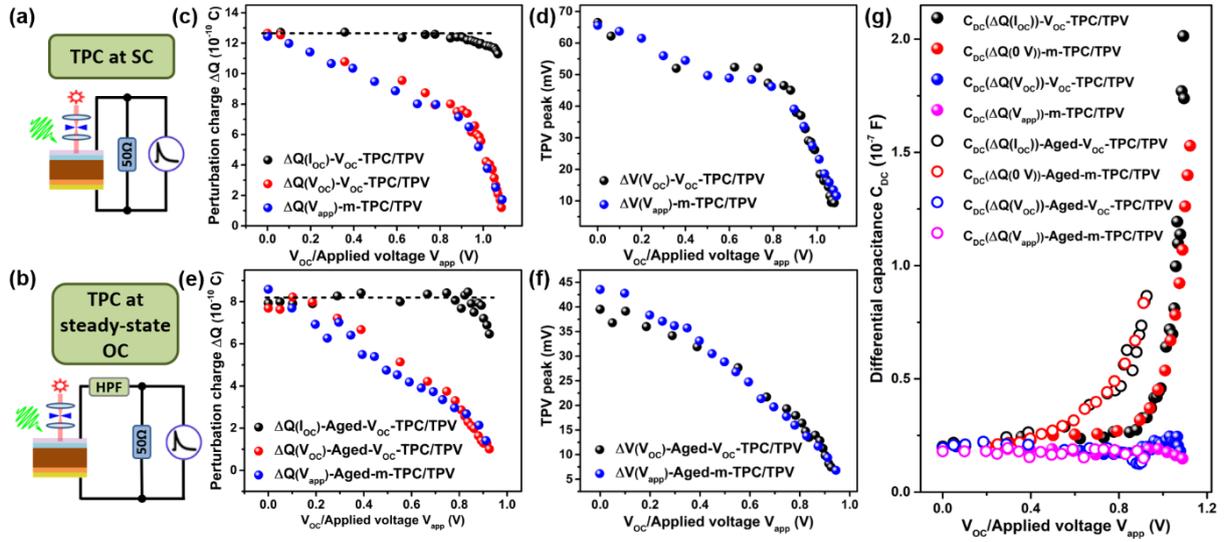

**Figure 4. Accurate measurement of the $C_{DC}$.** Schematic diagrams of perturbation charge measurement using (a) the conventional $V_{OC}$-TPC/TPV system or (b) a modified $V_{OC}$-TPC/TPV system by introducing an HPF which can help measure the TPC when the cell works under a steady-state OC condition. Integrated charge $\Delta Q (I_{OC})$ (black sphere), $\Delta Q (V_{OC})$ (red sphere) and $\Delta Q (V_{app})$ (blue sphere) for the $SnO_2$ and Aged-$SnO_2$ device, and comparisons of the TPV peak $\Delta V (V_{OC})$ (black sphere) and $\Delta V (V_{app})$ (red sphere) of these two cells (c. $\Delta Q$ and d. $\Delta V$ of the $SnO_2$ cell, e. $\Delta Q$ and f. $\Delta V$ of the Aged-$SnO_2$ cell). (g) $C_{DC}$ of these two cells computed from different methods (ESI). The conventional method gives an exponential behavior while a more accurate measurement of the perturbation charge yields a constant $C_{DC}$ across the whole voltage regime. No difference in the $C_{DC}$ can be observed between these two cells.

To obtain a deep understanding of the conventional scenario and the derived exponential $C_{DC}$-$V_{OC}$ relationship (Figure 3), we need to accurately measure the $\Delta Q (V_{OC})$ to check the validness of equation (1). By introducing a high-pass filter (HPF) in $V_{OC}$-TPC/TPV measurement system, we can measure the TPC signal even when the cell works in the steady-state OC condition, as schematically shown in Figure 4(a-b). The TPC of the standard and degraded cells at a series of $V_{OC}$ (Figure 4(b)) and $V_{app}$ (Figure 3(d)) were measured to obtain the $\Delta Q$ ($V_{OC}$) and $\Delta Q$ ($V_{app}$), respectively. The $\Delta Q$ ($I_{OC}$) (Figure 4(a)) is also presented in Figure 4(c) and (e) for a direct comparison. No matter the $V_{OC}$ or the $V_{app}$ mode, the $\Delta Q$ exhibit an obvious decrease at high voltage whereas the $\Delta Q$ ($I_{OC}$) is almost constant. This means that the widely used approximation $\Delta Q$ ($V_{OC}$) ≈ $\Delta Q$ ($I_{OC}$) is not valid in experiment, at least for our cells studied here. Further with the TPV peaks (i.e. $\Delta V$ ($V_{OC}$) and $\Delta V$ ($V_{app}$)) in Figure 4(d) and (f), the $C_{DC}$ of the cells is recalculated and shown in Figure 4(g). It can be seen that the exponential feature can only appear when the above inaccurate approximation is used. If without any artificial approximation, the $C_{DC}$ is almost constant in the whole voltage region, whatever the measurement mode. In this case, the analysis of tail state distribution through equation (3) is not valid any more. In addition, no difference in the $C_{DC}$ is observed for these two cells although they have exhibited obviously different carrier recombination characters. Thus, the applicability of the widely-used tail state framework for the cell electrical transient characterization, especially for the perovskite and other junction solar cells, need be reconsidered.

**Carrier dynamics and charge loss mechanism behind electrical transients**

To have a deep insight into the above clarification, two more fundamental questions about the electrical transient process of a solar cell will be unraveled here, that is, (1) how the charge loss within the cell influences the TPC and TPV, and (2) when and how the photovoltage is established.

The photo-generated charge ($\Delta Q_{tot}$) can be divided into four parts at the time point of TPV peak, that is,

$$\Delta Q_{tot} = \Delta Q_{ext}(V_{app}) + \Delta Q_R(V_{app}) + \Delta Q_{TS}(V_{app}) + \Delta Q_{CB/VB}(V_{app}), \qquad (4)$$

where $\Delta Q_{ext}$ is the extracted charge that located at the ETL and HTL/perovskite interfaces, $\Delta Q_R$ is the lost charge due to bulk recombination, $\Delta Q_{TS}$ is the charge that is located in the tail state, and $\Delta Q_{CB/VB}$ is the charge that still stored in the continuum conduction and valence band of the light absorber. It is reasonable to assume that the charge trapping and detrapping velocity of the tail state is in the similar order of magnitude. Thus, at a certain steady-state condition, the sum of $\Delta Q_{ext}$, $\Delta Q_{TS}$ and $\Delta Q_{CB/VB}$ can be estimated from the TPC integration. This sum is intrinsically determined by the charge loss item $\Delta Q_R (V_{app})$. In the conventional framework, this charge loss item has rarely been considered. This neglect may be applicable for the sensitized solar cell, because the charge transfer from dye (quantum dot) to the charge transporting layers is usually ultrafast;[57-60] whereas for the perovskite and other junction solar cells, it should be unreasonable because this charge transfer process usually costs tens to hundreds of nanosecond, in the similar time scale to that of bulk recombination.[16]

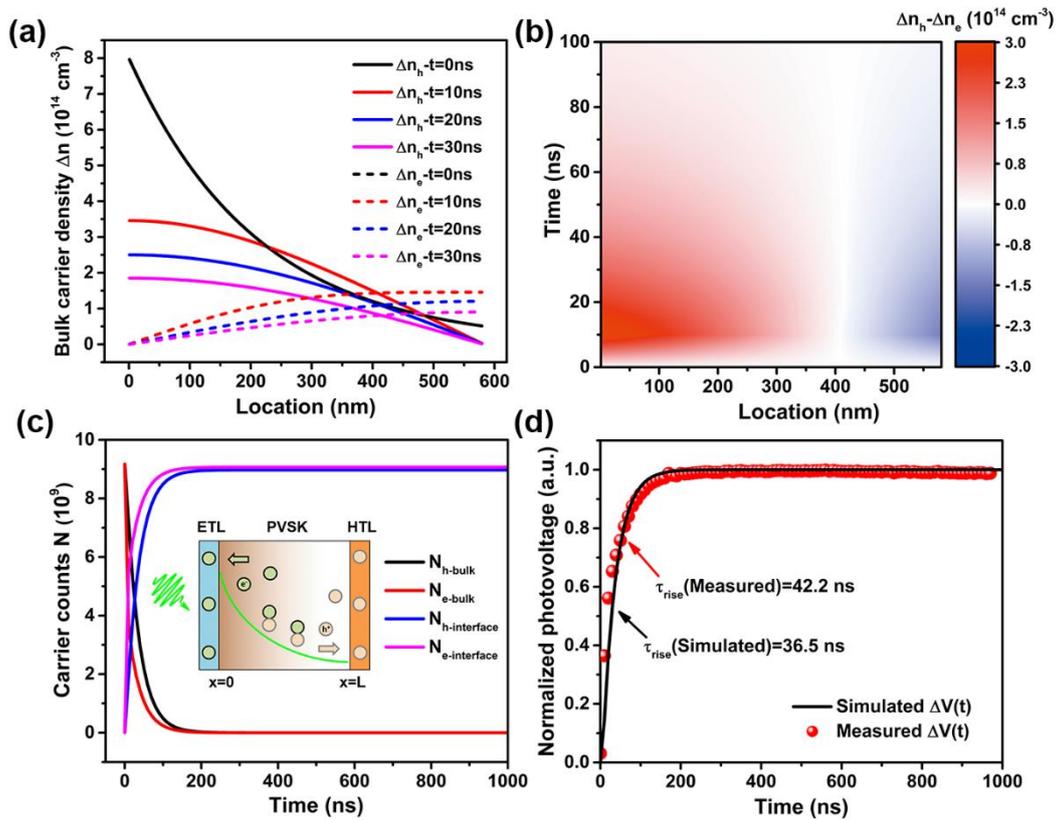

**Figure 5. Simulation of the charge transport within the perovskite absorber and the establishing process of the photovoltage.** (a) The distributions of photo-generated electron $n_e(x, t)$ and hole $n_h(x, t)$ within the perovskite absorber at several time points. (b) The contour plot of the net charge distribution within the perovskite absorber versus location and time. (c) The

time-dependent net charge within the perovskite and at the ETL and HTL interfaces. The inset is a schematic diagram of carrier transport processes within the diffusion model. (d) A simulation of the photovoltage rise process (black line) and a typical experimental result (red sphere) obtained from TPV.

We will further demonstrate that the peak photovoltage is established only after experiencing the charge transfer and bulk recombination processes. The photovoltage of the cell is simulated using Poisson's equation,[2] that is,

$$\frac{\partial^2 V(x,t)}{\partial x^2} = -\frac{\rho(x,t)}{\varepsilon}, \quad (5)$$

where the time-dependent charge distribution $\rho(x, t)$ (i.e. $q(n_h(x, t)-n_e(x, t))$. $n_h$: hole density, $n_e$: electron density, $x$: position, $t$: time) within the cell is obtained by considering the carrier transport and bulk recombination processes. Herein, the carrier drift behavior is ignored because the electric field within the perovskite solar cell has been significantly weakened by the ion migration.[61] Within the silicon solar cell, the width of charge neutral region (diffusion) is also much larger than that of the depletion region (drift). Due to the obvious difference in the carrier mobility between perovskite and the charge transporting layer, the extracted charge is considered locating at the interface with an approximated $\delta$ distribution with a total charge of $\Delta Q_{ext}$.

The electron and hole distributions within the perovskite absorber at several time points are presented in Figure 5(a). At the initial time, the electron and hole possess the same distribution and thus no net charge exists within the cell. With carrier transport, the electron and hole are separated in space due to their different boundary conditions. The resulted net charge distribution within the perovskite absorber is shown in Figure 5(b). The time-dependent net charge within the perovskite and at the ETL and HTL interfaces is further calculated and presented in Figure 5(c). After the charge transport and transfer for ~200 ns, all the net charges are located at the interface and charge loss due to the bulk recombination can be clearly seen.

According to equation (5) and Figure 5(b), the establishing behavior of the TPV is further derived by computing the time-dependent potential difference between the ETL and HTL (detail in ESI), that is, $\Delta V=|V_{ETL}-V_{HTL}|$. The theoretical result is given in Figure 5(d) with a solid line, where the TPV peak appears at ~200 ns. At this time point, net charge is no longer stored within the bulk perovskite but at the ETL and HTL interfaces, that is, $\Delta Q_{CB/VB}$ approaches zero and $\Delta Q_{ext}$ reaches its maximum of ~$(\Delta Q_{tot}-\Delta Q_R)$. For comparison, the TPV rising behavior of the cell is also measured and presented. Clearly, the theoretical result agrees well with the experimental result, demonstrating the reasonability of the above discussions. These results provide an unambiguous support for our first conclusion, that is, the charge loss arising from bulk recombination cannot be ignored when using the electrical transient characterizations. The decrease of $\Delta V$ ($V_{app}$) or $\Delta V$ ($V_{OC}$) at high voltages are mainly caused by the voltage enhanced bulk recombination, which cannot be reflected by the $C_{DC}$ feature because the $C_{DC}$ performs more like an electrode or geometry capacitance. Notably, if the tail state has a giant influence on the charge-carrier dynamics and distribution, the exponential $C_{DC}$ and $\Delta Q_{TS}$ features may be observed again since the charge located in the tail state could also influence the charge distribution and thus the TPV peak, although this phenomenon has never by been observed in our silicon and perovskite device experiments. For the sensitized or organic solar cell, this feature may be observed while is beyond the discussion of this work. Nonetheless, it is still an open question whether the transferred perturbation charge from the light absorber can immediately generate peak photovoltage within these cells.

**Quantification of charge loss and defect density by electrical transients**

As the conventional framework is invalid, a new way is urgently needed to quantify the charge loss and defect properties of the cell by using electrical transients. Here, for clarity, we deliberately separate the above charge-carrier dynamics into two processes, that is, (1) charge extraction at the perovskite interface and (2) charge collection at ETL and HTL/electrode interface. This dynamics separation is reasonable because they occur in an obviously distinct position and time scale. Extraction results from the carrier transport within the perovskite absorber (~ tens of nanosecond) while collection results from carrier transport within the ETL and HTL (~ several microseconds).[16] During the extraction, charge loss occurs through the bulk recombination mechanism; while for the collection, charge loses through the backward recombination at the ETL and HTL/perovskite interface. We have previously demonstrated that the charge collection efficiency ($\eta_C$) can be derived from the TPC ($\tau_C$) and TPV decay lifetime ($\tau_r$), that is,[16,33]

$$\eta_C (V) = 1 - \tau_C (V)/\tau_r (V). \tag{6}$$

Under a reasonable assumption that the internal quantum efficiency (IQE) of a cell at -1 V is approaching the unity, the IQE of the cell at different voltages can be estimated as

$$IQE (V) = \Delta Q (V)/\Delta Q (-1 \text{ V}), \tag{7}$$

where the $\Delta Q$ is obtained from the integration of TPC. The introduction of equation (7) can significantly improve the measurement accuracy compared to our previous route where the quantum efficiency was obtained by comparing the $\Delta Q$ and the perturbation light flux because there is giant inaccuracy in measuring the weak light pulse flux. Gathering the $\eta_C$ and IQE together, we can evaluate the charge extraction efficiency ($\eta_{ext}$) as,[33]

$$\eta_{ext} (V) = IQE (V)/\eta_C (V). \tag{8}$$

These three quantum efficiencies are used to quantify the charge loss within the cell and to discriminate the space-related charge loss mechanism. Simply, a low $\eta_{ext}(V)$ means a high charge recombination or inefficient charge transport within the light absorber, while a low $\eta_C$ (V) implies the interfacial backward recombination need be suppressed for higher device performance. Nonetheless, with the m-TPC/m-TPV and these extracted efficiencies, we can only give a simple qualitative description of the charge loss within the cell. This fact demands us to further investigate device physics of solar cells to establish a new analysis methodology correlated to the electrical transients.

Having a closer look at the charge processes within the bulk absorber, the $\eta_{ext}$ can be theoretically derived by considering the competition between charge transport and recombination within the absorber, that is,

$$\eta_{ext}(V) = \frac{\Delta Q_{ext}}{\Delta Q_{tot}} = \frac{\Delta Q_{ext}}{\Delta Q_{ext} + \Delta Q_R} = \frac{1}{1 + \tau_{ext}(V)/\tau_R(V)}, \qquad (9)$$

where $\tau_{ext}$ is the effective transport time of carriers within the bulk absorber, $\tau_R$ is the carrier recombination lifetime. According to the one-dimensional carrier diffusion model[62], $\tau_{ext}$ can be estimated from the carrier diffusion coefficient ($D$) and thickness of bulk absorber ($L$), as

$$\tau_{ext} = \frac{4}{D}\left(\frac{L}{\pi}\right)^2. \qquad (10)$$

This time parameter is almost independent to the bias voltage as long as the carrier transport mechanism within the bulk absorber has not been significantly changed. According to the semiconductor theory,[63] the carrier recombination within the absorber can be modulated by the bias voltage by changing the carrier density or the density of defect crossing with the Fermi energy level, which yields an approximate relationship of

$$\tau_R = \tau_{R0} e^{-\frac{qV}{AK_BT}} = \frac{1}{\sigma v_{th} N_t} e^{-\frac{qV}{AK_BT}}, \qquad (11)$$

where $\sigma$ is the charge trapping cross-section that can be estimated by a variety of experiments, $v_{th}$ is the thermal velocity calculated from the carrier effective mass, $N_t$ is the defect density and $A$ is a fitting factor that can be used to reflect the mechanism of bias voltage dependent carrier recombination. When $A$ is in the range between 1 and 2, the increase in charge recombination under higher voltage should arise from the increase in carrier density. When $A$ is larger than 2, we may need to consider the Fermi energy level dependent defect density. Gathering equation (10) and (11) into (9), we can then obtain the theoretical relationship between $\eta_{ext}$ and voltage, that is,

$$\eta_{ext} = \left[1 + \frac{4\sigma v_{th} N_t}{D}\left(\frac{L}{\pi}\right)^2 \exp\left(\frac{qV}{AK_BT}\right)\right]^{-1}, \qquad (12)$$

In the following, we will use these theoretical results to study the charge loss mechanism and defect density of the polycrystalline silicon, $Cu_2ZnSn(S, Se)_4$ (CZTS) and perovskite solar cells and to demonstrate the reliability of this new analysis method.

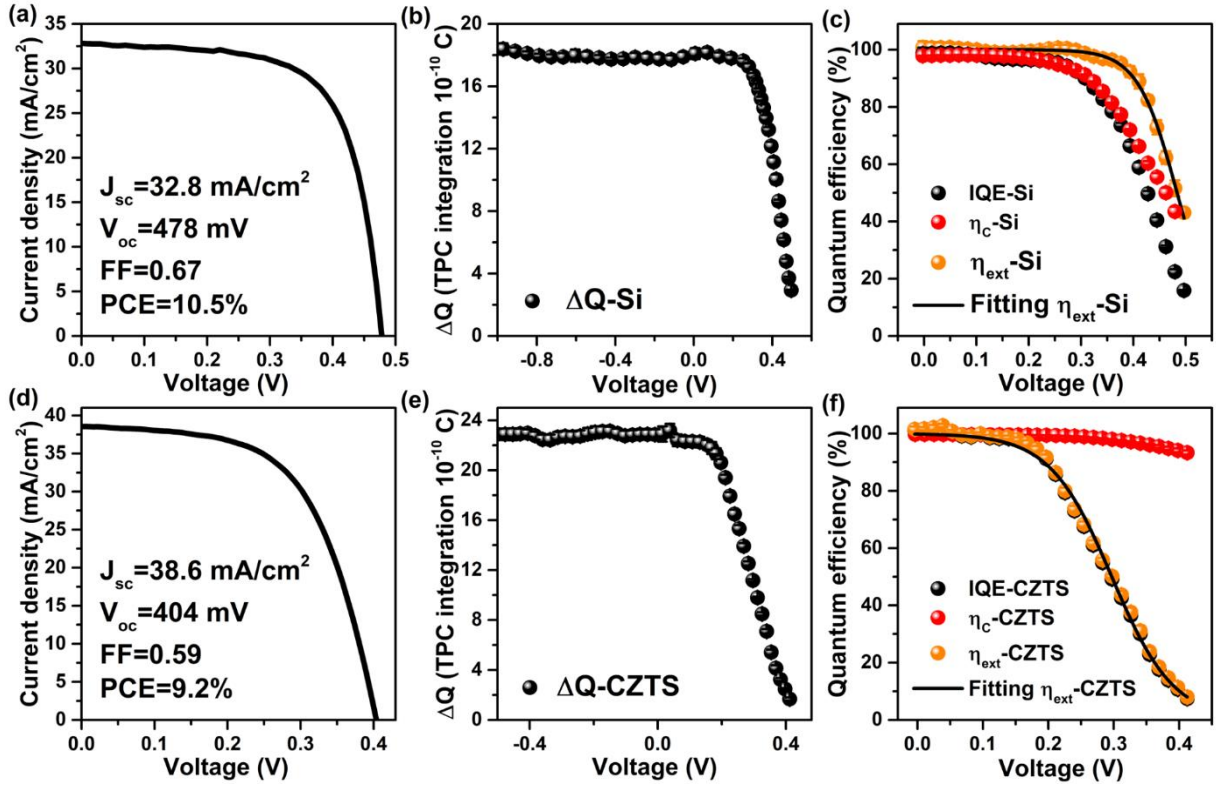

**Figure 6. Electrical transients study of a commercial polycrystalline silicon and lab-made CZTS solar cells.** (a, d) Representative *I-V* curves of the cells. (b, e) The relationship between the TPC integrated charge and the bias voltage. (c, f) The bias voltage dependent IQE (black sphere), $\eta_C$ (red sphere) and $\eta_{ext}$ (orange sphere). Black lines are the fitting of the $\eta_{ext}$ curve.

The application of this methodology to a silicon solar cell is presented in Figure 6(a-c). This commercial silicon solar cell was directly used and gives a PCE of 10.5% (Figure 6(a)). The TPC integration ($\Delta Q$ (V)) is firstly obtained at different voltages (Figure 6(b)), then used to derive the IQE (Figure 6(c)). The IQE exhibits a similar voltage-dependent behavior to that of the *I-V* curve. Using both TPC and TPV, the $\eta_C$ (V) is calculated, which decreases obviously when the voltage is higher than 0.3 V. Further, the $\eta_{ext}$ is derived from equation (8). From 0.3 to 0.45 V, the $\eta_{ext}$ is clearly larger than the $\eta_C$, which implies that the charge loss in this cell mainly occurs at the $n^+$-p silicon or the electrode interface when the charge transports within the $n^+$ layer before been collected. Usually, a surface or interface passivation by silicon oxides can suppress this interface backward recombination. According to equation

(12), the $\eta_{ext}$ (V) is fitted by using the previously reported $\sigma$ and $D$,[64-65] yielding $N_t$ of $4.75 \times 10^{12}$ cm$^{-3}$. This value is in the similar order of magnitude to the defect density of polycrystalline silicon material measured by other methods,[66-67] demonstrating the reliability of this electrical transient methodology proposed here.

For the CZTS solar cell, a different charge loss behavior is found, as presented in Figure 6(d-f). Its $\eta_C$ keeps higher than 90% across the studied voltage range while its $\eta_{ext}$ drops obviously at voltages higher than 0.2 V. This means that severe charge loss and low $V_{OC}$ issue that this cell encounters is mainly arisen from the bulk recombination. These results agree with recent findings that the CZTS absorber possesses complicated defect properties.[68] Since the $\eta_{ext}$ of the CZTS cell can be well described by our model, we infer that the width of built-in electric field is much smaller than the absorber thickness and the electric field is too weak to drive effective drift transport. Thus, for improving the CZTS performance, defect and heterojunction engineering is needed. The defect density is further estimated to be $2.1 \times 10^{15}$ cm$^{-3}$, also in a similar order of magnitude to that measured by other methods.[69]

**Table 1.** Device performance and derived defect density of the studied cells.

| Cell | $J_{SC}$ (mA cm$^{-2}$) | $V_{OC}$ (V) | FF | PCE (%) | $\sigma$ (cm$^2$) | $D$ (cm$^2$s$^{-1}$) | $A$ | $N_t$ (cm$^{-3}$) ET | $N_t$ (cm$^{-3}$) PL |
|---|---|---|---|---|---|---|---|---|---|
| Si | 32.8 | 0.48 | 0.67 | 10.5 | $6.0 \times 10^{-21(64)}$ | ~40[65] | 1.7 | $4.8 \times 10^{12}$ | - |
| CZTS | 38.6 | 0.40 | 0.59 | 9.2 | $3.5 \times 10^{-17(69)}$ | ~1.0[70] | 1.8 | $2.1 \times 10^{15}$ | - |
| P-A | 19.0 | 0.84 | 0.67 | 10.7 | $1.0 \times 10^{-17}$ | ~0.027 | 20.1 | $4.3 \times 10^{15}$ | $1.8 \times 10^{16}$ |
| P-B | 23.9 | 1.11 | 0.77 | 20.4 | $2.3 \times 10^{-16}$ | ~0.040 | 13.0 | $1.5 \times 10^{15}$ | $4.5 \times 10^{15}$ |

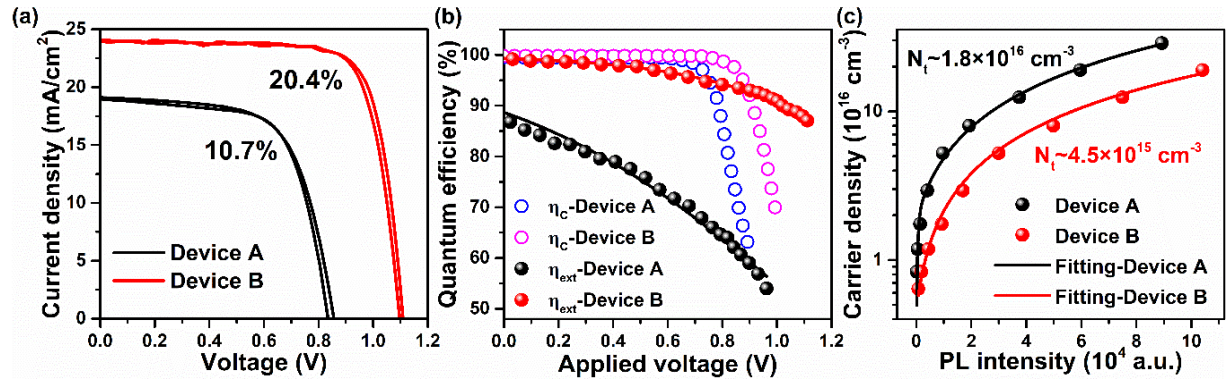

**Figure. 7 Electrical transients study of perovskite solar cells.** (a) The representative *I-V* curves of Device A and B, which were fabricated with different perovskite deposition conditions. (b) Voltage-dependent $\eta_C$ and $\eta_{ext}$ for these two devices. (c) The relationships between PL intensity and photoexcited carrier density for different perovskite films. The solid lines are numerical fits with the trap density model.

This method is further used to study the charge loss of perovskite solar cell. The perovskite absorber of Device A (P-A) is deposited by a conventional one-step spin coating method without any anti-solvent, which is supposed to possess a higher defect density and thus yields a low PCE of 10.7%. The perovskite absorber of Device B (P-B) is deposited by the widely-used anti-solvent method, contributing to a high PCE of 20.4%. For these two cells, $SnO_2$ nanoparticle film is used as the ETL and the hysteresis is negligible. The short-circuit current density ($J_{SC}$), $V_{OC}$ and fill factor (FF) of device A are all much lower than that of device B. The $\eta_C$ (V) and $\eta_{ext}$ (V) are derived from the m-TPC/TPV to discriminate the charge loss mechanism, especially that of Device A. As in Figure 7(b), these two cells possess a high $\eta_C$ approaching 100% at voltages lower than 0.8 V, which indicates that charge loss through the backward recombination at the ETL and HTL/perovskite interface in both devices is negligible at low voltages. This means that shunt contacts between the ETL and HTL due to the incomplete coverage of the perovskite film that may appear in the conventional spin coting method does not dominate the charge loss of the cells here. For Device A, its $\eta_C$ begins to decrease obviously from 0.8 V while $\eta_C$ of Device B begins to decrease at ~0.9 V. This difference in the $\eta_C$ could result from the different built-in potential ($V_{bi}$) of these two devices since a weakening of built-in electric field would significantly increase the interface charge loss through backward recombination. For Device B, the $\eta_C$ at high voltages need to be further improved to boost its current $V_{OC}$, which is mainly determined by the interface energy level structures and charge recombination. It is suggested that this is the key mechanism for the charge loss of the currently-reported highest efficient perovskite solar cells.

The origin for the device performance difference mainly lies in the $\eta_{ext}$ (V). In the SC condition, the $\eta_{ext}$ for device A is only ~86% while device B possesses a high $\eta_{ext}$ of 99%. This makes a great contribution to the $J_{SC}$ difference and indicates that charge loss for device A mainly occurs in the perovskite absorber. Further according to equation (12) and the experimentally measured $\sigma$, $D$ and $L$, the defect density of the perovskite absorber in Device A is estimated to be ~$4.3 \times 10^{15}$ cm$^{-3}$, which is about three times as high as that of Device B. These values obtained by the electrical transient method (ET) are approximately two orders of magnitude lower than that measured from thermal admittance spectroscopy.[71-72] This may arise because not all the defects probed by the capacitance response could make contribution to the bulk charge recombination. Interface defects induced by the ion migration may also influence the accurate measurement of bulk defect. For a more reasonable comparison, we also used excitation intensity dependent photoluminescence to estimate the perovskite defect density, as presented in Figure 7(c) and Table 1.[26] The result is in the similar order of magnitude to that obtained by the ET method.

More interestingly, we find that the fitted $A$ for the perovskite solar cell is much higher than 2, indicating that the bulk recombination is not sensitive to the bias voltage. This may arise from two possible reasons, that is, (1) ion migration has screened the influence of bias voltage on the bulk perovskite[73] or (2) the distinctive defect distribution of the perovskite has influenced the dependence of carrier recombination lifetime to the bias voltage.[74] To check these possibilities, the defect analysis is further carried out on MAPbBr$_3$ based cell because these two types of perovskite should possess similar defect distribution properties but significantly different ion migration behaviors. It is found that the MAPbBr$_3$ cell exhibits a much smaller $A$ of ~5.2 (ESI). This result helps us to attribute the large $A$ observed for Device A and B to the screening effect of ion migration.

## 3. Conclusions

In summary, this work has comprehensively revealed how to exploit electrical transients for a quantitative study of charge loss within photovoltaic devices. A more accurate evaluation of voltage-dependent differential capacitances together with a clear description of the photovoltage establishing process is put forward to verify the unreasonability of the previously widely-used tail-state framework. Alternatively, we provide a new way to quantifying charge loss and defect density of solar cells by electrical transients. Quantum efficiency of charge transfer at different interfaces (i.e. $\eta_C$ and $\eta_{ext}$) can be extracted to reveal the charge loss mechanism of a solar cell; moreover, the defect density of the light absorber is also measured from the $\eta_{ext}$-voltage behaviors. It has been successfully applied to the polycrystalline silicon, CZTS and perovskite solar cells to confirm its reliability. More importantly, this methodology is supposed to be applicable to other semiconductor energy devices with similar structures, such as the CIGS and $Sb_2Se_3$ planar junction solar cells. Therefore, our work provides a more accurate understanding toward dynamic physics processes and charge loss mechanisms of solar cells and other related semiconductor energy devices by the aid of electrical transient techniques. In the meantime, it also makes a great step forward to exploring the essential properties of junction solar cells, giving inspirations for solving the crucial issues and having values for developing other semiconductor materials and devices.

**Experimental Section**

**Materials**

$PbI_2$ (99.9985%) and $SnO_2$ colloidal dispersion (tin (IV) oxide, 15% in $H_2O$ colloidal dispersion) were obtained from Alfa Asear. $PbBr_2$ (99.999%) were obtained from Sigma-Aldrich. Methylammonium bromide (MABr), methylammonium iodide (MAI) and formamidinium iodide (FAI) were obtained from Xi'an Polymer Light Technology Corp. N,N-Dimethylformamide (DMF), dimethylsulfoxide (DMSO), chlorobenzene (CB) and tert-butylpyridine (TBP, 99.999%) were obtained from Alfa Aesar. Spiro-OMeTAD was obtained from Luminescence Technology Corp. (Lumtec). Bis(trifluoromethane)sulfonimide lithium salt (LiTFSI) was purchased from Sigma-Aldrich (99.9985%), and

tris(2-(1H-pyrazol-1-yl)-4-tert-butylpyridine)-cobalt(III) tris(bis(trifluorom-ethylsulfonyl)imide) (FK209) was from Dyesol.

**Device fabrication**

Laser-patterned ITO glass (sheet resistance= 15 Ω per square) was pre-cleaned in an ultrasonic bath using soap water, deionized water, acetone and ethanol for 30 min each, and treated in an ultraviolet–ozone chamber for 15 min. A thin layer of $SnO_2$ nanoparticles was spin coated onto the pre-cleaned ITO substrate using a $SnO_2$ colloidal dispersion solution in water (6.18%) at 5000 rpm for 30 s and annealed at 150 ℃ for 30 min in ambient atmosphere. [6,6]-phenyl-C61-butyric acid (PCBA, 0.1 mg/mL) in CB was spin-coated on the top of $SnO_2$ film.[75] The precursor solution for perovskite films was prepared by dissolving $PbI_2$ (1.32 M), $PbBr_2$ (0.12 M), FAI (1.08 M), MAI (0.12 M) and MABr (0.24 M) in DMF/DMSO mixed solvent (v:v= 4:1) and stirred overnight at room temperature. The perovskite film was spin coated by an anti-solvent one-step method in the glove box. The perovskite precursor solution was spin-coated at 1000 rpm for 10 s and 5000 rpm for 30 s, CB (120 μL) was poured onto the substrate at 15 s during the high speed stage. Then the half-crystallization film was heated at 150 ℃ in the glove box for 10 min and 100 ℃ in vacuum for 40 min. Then, a 200 nm-thick spiro-OMeTAD layer was deposited by spin-coating at 3500 rpm for 30 s and then heated for 10 min at 60 ℃ on the top of perovskite films. The spiro-OMeTAD solution was prepared by dissolving in CB (60 mM) with the additives of Li-TFSI, FK209 and TBP at doping molar ratios of 0.5, 0.03 and 3.3. Finally, 80 nm thick Au electrodes were deposited via thermal evaporation at an atmospheric pressure of $10^{-7}$ Torr (Kurt J. Lesker). The multicrystalline silicon solar cell was directly purchased from Stark Electronics, China. The CZTS solar cells were fabricated based on our previous work.[76]

**Characterization**

J-V characteristics were measured under AM 1.5 simulated sunlight (100 mW/cm$^2$) from Zolix SS150A, which was recorded by a digital source meter (Keithley model 2602).

External quantum efficiency (EQE) of the devices was measured with a lab-made setup under 0.3-0.9 mW/cm$^2$ monochromic light illumination. The wavelength was from 300 nm to 850 nm. Light intensity-dependent $I_{sc}$ and $V_{oc}$ curves were performed by the lab-made instrument mainly using an intensity tunable light emitting diode (Nanjin Hongzhao, S3000) and an IM6ex electrochemical workstation (Zahner). Modulated transient photocurrent and photovoltage (m-TPC/TPV) measurements were obtained by a tunable nanosecond laser (Opotek, RADIANT 532 LD) pumped at 532 nm and recorded by a sub-nanosecond resolved digital oscilloscope (Tektronix, MDO3034) with input impedances of 50 Ω or 1 MΩ, respectiely. A signal generator (Tektronix, AFG3052C) together with a low-pass filter was applied to give steady-state bias voltages over the cell. Traditional $V_{oc}$-TPC/TPV measurements were obtained by a tunable nanosecond laser (Opotek, RADIANT 532 LD) pumped at 532 nm and recorded by a sub-nanosecond resolved digital oscilloscope (Tektronix, MDO3034) with input impedances of 50 Ω or 1 MΩ, respectiely. The background illumination was provided by an intensity tunable light emitting diode (Nanjin Hongzhao, S3000). The intensity was first calibrated to the $J_{sc}$ of devices measured under AM1.5 solar simulator. Time-resolved photoluminescence (TRPL) was measured by the PL spectrometer (Edinburgh Instruments, FLS900) using a pulsed diode laser (EPL-445, ~20 nJ cm$^{-2}$/pulse) as the excitation source. The thickness of the perovskite films was measured by the cross-sectional SEM image as well as a surface profiler (KLA-Tencor). Thermal admittance spectroscopy (TAS) was performed using an IM6ex electrochemical workstation (Zahner), in which the scanning frequency was set between 0.1 and 10$^5$ Hz, and the amplitude of the sine perturbation bias was set to 10 mV.

## AUTHOR INFORMATION

**Corresponding Author**

qbmeng@iphy.ac.cn

**Notes**

The authors declare no competing financial interests.

ACKNOWLEDGMENT

This work was supported by Natural Science Foundation of China (Nos. 51421002, 51627803, 91733301, 51761145042, 53872321, 51572288 and 11874402), the International Partnership Program of Chinese Academy of Sciences (No. 112111KYSB20170089) and the National Key R&D Program of China (2018YFB1500101).

**REFERENCES**

1. S. G. Benka, The energy challenge, *Phys. Today*, 2002, **55**, 38-39.
2. A. Luque and S. Hegedus, Handbook of photovoltaic science and engineering, *John Wiley & Sons*, Hoboken, New Jersey, 2011.
3. P. R. Gray, P. J. Hurst, S. H. Lewis and R. G. Meyer, Analysis and Design of Analog Integrated Circuits, *John Wiley & Sons*, Hoboken, New Jersey, 2009.
4. J. M. Rabaey, Digital Integrated Circuits, *Prentice-Hall*, Englewood Cliffs, 1996.
5. Y. Yang, D. P. Ostrowski, R. M. France, K. Zhu, J. van de Lagemaat, J. M. Luther and M. C. Beard, Observation of a hot-phonon bottleneck in lead-iodide perovskites, *Nat. Photon.*, 2015, **10**, 53-59.
6. J. Fu, Q. Xu, G. Han, B. Wu, C. H. A. Huan, M. L. Leek and T. C. Sum, Hot carrier cooling mechanisms in halide perovskites, *Nat. Commun.*, 2017, **8**, 1300.
7. H. Zhu, K. Miyata, Y. Fu, J. Wang, P. P. Joshi, D. Niesner, K. W. Williams, S. Jin and X. Y. Zhu, Screening in crystalline liquids protects energetic carriers in hybrid perovskites, *Science*, 2016, **353**, 1409-1413.
8. J. Yang, X. Wen, H. Xia, R. Sheng, Q. Ma, J. Kim, P. Tapping, T. Harada, T. W. Kee, F. Huang, Y. Cheng, M. Green, A. Ho-Baillie, S. Huang, S. Shrestha, R. Patterson and G. Conibeer, Acoustic-optical phonon up-conversion and hot-phonon bottleneck in lead-halide perovskites, *Nat. Commun.*, 2017, **8**, 14120.
9. T. Leijtens, G. E. Eperon, N. K. Noel, S. N. Habisreutinger, A. Petrozza and H. J. Snaith, Stability of Metal Halide Perovskite Solar Cells, *Adv. Energy Mater.*, 2015, **5**, 1500963.
10. S. Meloni, T. Moehl, W. Tress, M. Franckevičius, M. Saliba, Y. H. Lee, P. Gao, M. K. Nazeeruddin, S. M. Zakeeruddin, U. Rothlisberger and M. Grätzel, Ionic polarization-induced current–voltage hysteresis in $CH_3NH_3PbX_3$ perovskite solar cells, *Nat. Commun.*, 2016, **7**, 10334.
11. B. Chen, X. Zheng, M. Yang, Y. Zhou, S. Kundu, J. Shi, K. Zhu and S. Priya, Interface band structure engineering by ferroelectric polarization in perovskite solar cells, *Nano Energy*, 2015, **13**, 582-591.

12. C. Li, S. Tscheuschner, F. Paulus, P. E. Hopkinson, J. Kießling, A. Köhler, Y. Vaynzof and S. Huettner, Iodine Migration and its Effect on Hysteresis in Perovskite Solar Cells, *Adv. Mater.*, 2016, **28**, 2446-2454.
13. Y. Shao, Y. Fang, T. Li, Q. Wang, Q. Dong, Y. Deng, Y. Yuan, H. Wei, M. Wang, A. Gruverman, J. Shield and J. Huang, Grain boundary dominated ion migration in polycrystalline organic-inorganic halide perovskite films, *Energy Environ. Sci.*, 2016, **9**, 1752-1759.
14. K. Domanski, B. Roose, T. Matsui, M. Saliba, S. H. Turren-Cruz, J. P. Correa-Baena, C. R. Carmona, G. Richardson, J. M. Foster, F. De Angelis, J. M. Ball, A. Petrozza, N. Mine, M. K. Nazeeruddin, W. Tress, M. Grätzel, U. Steiner, A. Hagfeldt and A. Abate, Migration of cations induces reversible performance losses over daynight cycling in perovskite solar cells, *Energy Environ. Sci.*, 2017, **10**, 604-613.
15. J. P. Correa-Baena, M. Saliba, T. Buonassisi, M. Grätzel, A. Abate, W. Tress and A. Hagfeldt, Promises and challenges of perovskite solar cells, *Science*, 2017, **358**, 739-744.
16. J. Shi, Y. Li, Y. Li, D. Li, Y. Luo, H. Wu and Q. Meng, From Ultrafast to Ultraslow Charge-Carrier Dynamics of Perovskite Solar Cells, *Joule*, 2018, **2**, 879-901.
17. Z. Zhu, J. Ma, Z. Wang, C. Mu, Z. Fan, L. Du, Y. Bai, L. Fan, H. Yan, D. L. Phillips and S. Yang, Efficiency enhancement of perovskite solar cells through fast electron extraction: the role of graphene quantum dots, *J. Am. Chem. Soc.*, 2014, **136**, 3760-3763.
18. Z. Guo, Y. Wan, M. Yang, J. Snaider, K. Zhu and L. Huang, Long-range hot-carrier transport in hybrid perovskites visualized by ultrafast microscopy, *Science*, 2017, **356**, 59-62.
19. P. Piatkowski, B. Cohen, C. S. Ponseca Jr., M. Salado, S. Kazim, S. Ahmad, V. Sundström and A. Douhal, Unraveling charge carriers generation, diffusion, and recombination in formamidinium lead triiodide perovskite polycrystalline thin film, *J. Phys. Chem. Lett.*, 2016, **7**, 204-210.
20. D. A. Valverde-Chávez, C. S. Ponseca Jr., C. C. Stoumpos, A. Yartsev, M. G. Kanatzidis, V. Sundström and D. G. Cooke, Intrinsic femtosecond charge generation dynamics in single crystal $CH_3NH_3PbI_3$, *Energy Environ. Sci.*, 2015, **8**, 3700-3707.
21. C. Wehrenfennig, M. Liu, H. J. Snaith, M. B. Johnston and L. M. Herz, Charge-carrier dynamics in vapour-deposited films of the organolead halide perovskite $CH_3NH_3PbI_{3-x}Cl_x$, *Energy Environ. Sci.*, 2014, **7**, 2269-2275.
22. G. Xing, N. Mathews, S. Sun, S. Lim, Y. M. Lam, M. Grätzel, S. Mhaisalkar and T. C. Sum, Long-Range Balanced Electron- and Hole-Transport Lengths in Organic-Inorganic $CH_3NH_3PbI_3$, *Science*, 2013, **342**, 344-347.
23. S. D. Stranks, G. E. Eperon, G. Grancini, C. Menelaou, M. J. Alcocer, T. Leijtens, L. M. Herz, A. Petrozza and H. J. Snaith, Electron-Hole Diffusion Lengths Exceeding 1 Micrometer in an Organometal Trihalide Perovskite Absorber, *Science*, 2013, **342**, 341-344.
24. S. D. Stranks, V. M. Burlakov, T. Leijtens, J. M. Ball, A. Goriely and H. J. Snaith, Recombination kinetics in organic-inorganic perovskites excitons, free charge, and subgap states, *Phys. Rev. Appl.*, 2014, **2**, 034007.

25 Y. Yamada, T. Nakamura, M. Endo, A. Wakamiya and Y. Kanemitsu, Photocarrier Recombination Dynamics in Perovskite $CH_3NH_3PbI_3$ for Solar Cell Applications, *J. Am. Chem. Soc.*, 2014, **136**, 11610-11613.

26 Y. Li, Y. Li, J. Shi, H. Zhang, J. Wu, D. Li, Y. Luo, H. Wu and Q. Meng, High Quality Perovskite Crystals for Efficient Film Photodetectors Induced by Hydrolytic Insulating Oxide Substrates, *Adv. Funct. Mater.*, 2018, **28**, 1705220.

27 C. S. Ponseca Jr., T. J. Savenije, M. Abdellah, K. Zheng, A. Yartsev, T. Pascher, T. Harlang, P. Chabera, T. Pullerits, A. Stepanov, J. P. Wolf and V. Sundström, Organometal Halide Perovskite Solar Cell Materials Rationalized: Ultrafast Charge Generation, High and Microsecond-Long Balanced Mobilities, and Slow Recombination, *J. Am. Chem. Soc.*, 2014, **136**, 5189-5192.

28 H. Oga, A. Saeki, Y. Ogomi, S. Hayase and S. Seki. Improved Understanding of the Electronic and Energetic Landscapes of Perovskite Solar Cells High Local Charge Carrier Mobility, Reduced Recombination, and Extremely Shallow Traps, *J. Am. Chem. Soc.*, 2014, **136**, 13818-13825.

29 D. Guo, V. M. Caselli, E. M. Hutter and T. J. Savenije, Comparing the Calculated Fermi Level Splitting with the Open-Circuit Voltage in Various Perovskite Cells, *ACS Energy Lett.*, 2019, **4**, 855-860.

30 D. Kiermasch, A. Baumann, M. Fischer, V. Dyakonov and K. Tvingstedt, Revisiting lifetimes from transient electrical characterization of thin film solar cells; a capacitive concern evaluated for silicon, organic and perovskite devices, *Energy Environ. Sci.*, 2018, **11**, 629-640.

31 T. Du, J. Kim, J. Ngiam, S. Xu, P. R. F. Barnes, J. R. Durrant and M. A. McLachlan, Elucidating the Origins of Subgap Tail States and Open Circuit Voltage in Methylammonium Lead Triiodide Perovskite Solar Cells, *Adv. Funct. Mater.*, 2018, **28**, 1801808.

32 S. Wheeler, D. Bryant, J. Troughton, T. Kirchartz, T. Watson, J. Nelson and J. R. Durrant, Transient Optoelectronic Analysis of the Impact of Material Energetics and Recombination Kinetics on the Open-Circuit Voltage of Hybrid Perovskite Solar Cells, *J. Phys. Chem. C*, 2017, **121**, 13496-13506.

33 J. Shi, D. Li, Y. Luo, H. Wu and Q. Meng. Opto-electro-modulated transient photovoltage and photocurrent system for investigation of charge transport and recombination in solar cells, *Rev. Sci. Instrum.*, 2016, **87**, 123107.

34 J. Shi, H. Zhang, X. Xu, D. Li, Y. Luo and Q. Meng. Microscopic Charge Transport and Recombination Processes behind the Photoelectric Hysteresis in Perovskite Solar Cells, *Small*, 2016, **12**, 5288-5294.

35 J. Shi, X. Xu, H. Zhang, Y. Luo, D. Li and Q. Meng, Intrinsic slow charge response in the perovskite solar cells: Electron and ion transport, *Appl. Phys. Lett.*, 2015, **107**, 163901.

36 S. Wood, J. C. Blakesley and F. A. Castro, Assessing the Validity of Transient Photovoltage Measurements and Analysis for Organic Solar Cells, *Phys. Rev. Applied*, 2018, **10**, 024038.

37  O. J. Sandberg, K. Tvingstedt, P. Meredith and A. Armin, Theoretical Perspective on Transient Photovoltage and Charge Extraction Techniques, *J. Phys. Chem. C*, 2019, **123**, 14261-14271.
38  M. O. Giulio, S. Galassini, G. Micocci, A. Tepore and C. Manfredotti, Determination of minority-carrier lifetime in silicon solar cells from laser-transient photovoltaic effect, *J. Appl. Phys.*, 1981, **52**, 7219-7223.
39  Y. Sang, W. Liu, F. Qiao, D. Kang and A. Liu, Transient photovoltage in poly(3-hexylthiophene)n-crystalline-silicon heterojunction, *Vacuum*, 2013, **93**, 28-30.
40  X. Wang, S. Karanjit, L. Zhang, H. Fong, Q. Qiao and Z. Zhu, Transient photocurrent and photovoltage studies on charge transport in dye sensitized solar cells made from the composites of $TiO_2$ nanofibers and nanoparticles, *Appl. Phys. Lett.*, 2011, **98**, 082114.
41  E. Enache-Pommer, J. E. Boercker and E. S. Aydil, Electron transport and recombination in polycrystalline $TiO_2$ nanowire dye-sensitized solar cells, *Appl. Phys. Lett.*, 2007, **91**, 123116.
42  S. Pang, K. Cheng, Z. Yuan, S. Xu, G. Cheng and Z. Du, Study on dynamics of photoexcited charge injection and trapping in CdS quantum dots sensitized $TiO_2$ nanowire array film electrodes, *Appl. Phys. Lett.*, 2014, **104**, 201601.
43  C. Z. Li, C. Y. Chang, Y. Zang, H. X. Ju, C. C. Chueh, P. W. Liang, N. Cho, D. S. Ginger and A. K. Y. Jen, Suppressed Charge Recombination in Inverted Organic Photovoltaics via Enhanced Charge Extraction by Using a Conductive Fullerene Electron Transport Layer, *Adv. Mater.*, 2014, **26**, 6262-6267.
44  B. C. O'Regan, S. Scully, A. C. Mayer, E. Palomares and J. R. Durrant, The Effect of $Al_2O_3$ Barrier Layers in $TiO_2$/Dye/CuSCN Photovoltaic Cells Explored by Recombination and DOS Characterization Using Transient Photovoltage Measurements, *J. Phys. Chem. B*, 2005, **109**, 4616-4623.
45  C. G. Shuttle, B. C. O'Regan, A. M. Ballantyne, J. Nelson, D. D. C. Bradley, J. de Mello and J. R. Durrant, Experimental determination of the rate law for charge carrier decay in a polythiophene Fullerene solar cell, *Appl. Phys. Lett.*, 2008, **92**, 093311.
46  A. Maurano, C. G. Shuttle, R. Hamilton, A. M. Ballantyne, J. Nelson, W. Zhang, M. Heeney and J. R. Durrant, Transient Optoelectronic Analysis of Charge Carrier Losses in a Selenophene/Fullerene Blend Solar Cell, *J. Phys. Chem. C*, 2011, **115**, 5947-5957.
47  F. Marlow, A. Hullermann and L. Messmer. Is the Charge Transport in Dye-Sensitized Solar Cells Really Understood?, *Adv. Mater.*, 2015, **27**, 2447-2452.
48  Y. Shao, Y. Yuan and J. Huang, Correlation of energy disorder and open-circuit voltage in hybrid perovskite solar cells, *Nat. Energy*, 2016, **1**, 15001.
49  A. Zohar, M. Kulbak, I. Levine, G. Hodes, A. Kahn and D. Cahen, What Limits the Open-Circuit Voltage of Bromide Perovskite-Based Solar Cells?, *ACS Energy Lett.*, 2019, **4**, 1-7.
50  A. D. Wright, R. L. Milot, G. E. Eperon, H. J. Snaith, M. B. Johnston and L. M. Herz, Band‐Tail Recombination in Hybrid Lead Iodide Perovskite, *Adv. Funct. Mater.*, 2017, **27**, 1700860.

51 J. Wu, J. Shi, Y. Li, H. Li, H. Wu, Y. Luo, Dongmei Li and Qingbo Meng, Quantifying the Interface Defect for the Stability Origin of Perovskite Solar Cells, *Adv. Energy Mater.*, 2019, **9**, 1901352.

52 Y. Li, Y. Li, J. Shi, H. Li, H. Zhang, J. Wu, D. Li, Y. Luo, H. Wu and Qingbo Meng, Photocharge accumulation and recombination in perovskite solar cells regarding device performance and stability, *Appl. Phys. Lett.*, 2018, **112**, 053904.

53 M. Stolterfoht, P. Caprioglio, C. M. Wolff, J. A. Márquez, J. Nordmann, S. Zhang, D. Rothhardt, U. Hörmann, Y. Amir, A. Redinger, L. Kegelmann, F. Zu, S. Albrecht, N. Koch, T. Kirchartz, M. Saliba, T. Unold and D. Neher, The impact of energy alignment and interfacial recombination on the internal and external open-circuit voltage of perovskite solar cells, *Energy Environ. Sci.*, 2019, **12**, 2778-2788.

54 L. T. Schelhas, Z. Li, J. A. Christians, A. Goyal, P. Kairys, S. P. Harvey, D. H. Kim, K. H. Stone, J. M. Luther, Kai Zhu, V. Stevanovice and J. J. Berry, Insights into operational stability and processing of halide perovskite active layers, *Energy Environ. Sci.*, 2019, **12**, 1341-1348.

55 X. Liu, X. Du, J. Wang, C. Duan, X. Tang, T. Heumueller, G. Liu, Y. Li, Z. Wang, J. Wang, F. Liu, N. Li, C. J. Brabec, F. Huang and Y. Cao, Efficient Organic Solar Cells with Extremely High Open-Circuit Voltages and Low Voltage Losses by Suppressing Nonradiative Recombination Losses, *Adv. Energy Mater.*, 2018, **8**, 1801699.

56 W. Tress, M. Yavari, K. Domanski, P. Yadav, B. Niesen, J. P. C. Baena, A. Hagfeldt and M. Grätzel, Interpretation and evolution of open-circuit voltage, recombination, ideality factor and subgap defect states during reversible light-soaking and irreversible degradation of perovskite solar cells, *Energy Environ. Sci.*, 2018, **11**, 151-165.

57 J. Bisquert, A. Zaban, M. Greenshtein and I. Mora-Sero, Determination of Rate Constants for Charge Transfer and the Distribution of Semiconductor and Electrolyte Electronic Energy Levels in Dye-Sensitized Solar Cells by Open-Circuit Photovoltage Decay Method, *J. Am. Chem. Soc.*, 2004, **126**, 13550-13559.

58 E. Martínez-Ferrero, I. M. Seró, J. Albero, S. Giménez, J. Bisquert and E. Palomares, Charge transfer kinetics in CdSe quantum dot sensitized solar cells, *Phys. Chem. Chem. Phys.*, 2010, **12**, 2819-2821.

59 M. Pazoki, U. B. Cappel, E. M. J. Johansson, A. Hagfeldt and G. Boschloo, Characterization techniques for dye-sensitized solar cells, *Energy Environ. Sci.*, 2017, **10**, 672-709.

60 A. Listorti, B. C. O'Regan and J. R. Durrant, Electron Transfer Dynamics in Dye-Sensitized Solar Cells, *Chem. Mater.*, 2011, **23**, 3381-3399.

61 N. E. Courtier, J. M. Cave, J. M. Foster, A. B. Walker and G. Richardson, How transport layer properties affect perovskite solar cell performance insights from a coupled charge transport/ion migration model, *Energy Environ. Sci.*, 2019, **12**, 396-409.

62 A. B. Sproul, Dimensionless solution of the equation describing the effect of surface recombination on carrier decay in semiconductors, *J. Appl. Phys.*, 1994, **76**, 2851-2854.

63 S. M. Sze and K. K. Ng, Physics of Semiconductor Devices (3$^{rd}$ Edition), *John Wiley & Sons*, Hoboken, New Jersey, 2007.

64  A. Criado, B, Alonso and J. Piqueras, Deep traps in polysilicon solar cells, *Electron. Lett.*, 1978, **14**, 622-623.

65  J. Martinez, B. Alonso, A. Criado and J. Piqueras, Solar cells from polysilicon rods, *Electron. Lett.*, 1976, **12**, 671-672.

66  A. Datta, J. Damon-Lacoste, P. R. i Cabarrocas and P. Chatterjee, Defect states on the surfaces of a P-type c-Si wafer and how they control the performance of a double heterojunction solar cell, *Sol. Energy Mat. Sol. Cells*, 2008, **92**, 1500-1507.

67  Y. Park, J. Lu, J. H. Park and G. Rozgonyi, Impact of Structural Defect Density on Gettering of Transition Metal Impurities during Phosphorus Emitter Diffusion in Multi-Crystalline Silicon Solar Cell Processing, *Electron. Mater. Lett.*, 2015, **11**, 658-663.

68  K. Pal, P. Singh, A. Bhaduri and K. B.Thapa, Current challenges and future prospects for a highly efficient (>20%) kesterite CZTS solar cell: A review, *Sol. Energ. Mat. Sol. C.*, 2019, **196**, 138-156.

69  H. S. Duan, W. Yang, B. Bob, C. J. Hsu, B. Lei and Y. Yang, The Role of Sulfur in Solution-Processed $Cu_2ZnSn(S,Se)_4$ and its Effect on Defect Properties, *Adv. Funct. Mater.*, 2013, **23**, 1466-1471.

70  T. Gokmen, O. Gunawan and D. B. Mitzi, Minority carrier diffusion length extraction in $Cu_2ZnSn(Se,S)_4$ solar cells, *J. Appl. Phys.*, 2013, **114**, 114511.

71  G. Xing, N. Mathews, S. S. Lim, N. Yantara, X. Liu, D. Sabba, M. Grätzel, S. Mhaisalkar and T. C. Sum, Low-temperature solution-processed wavelength-tunable perovskites for lasing, *Nat. Mater.*, 2014, **13**, 476-480.

72  V. Adinolfi, W. Peng, G. Walters, O. M. Bakr and E. H. Sargent, The Electrical and Optical Properties of Organometal Halide Perovskites Relevant to Optoelectronic Performance, *Adv. Mater.*, 2018, **30**, 1700764.

73  R. A. Belisle, W. H. Nguyen, A. R. Bowring, P. Calado, X. Li, S. J. C. Irvine, M. D. McGehee, P. R. F. Barnes and B. C. O'Regan, Interpretation of inverted photocurrent transients in organic lead halide perovskite solar cells: proof of the field screening by mobile ions and determination of the space charge layer widths, *Energy Environ. Sci.*, 2017, **10**, 192-204.

74  J. Shi, H. Zhang, Y. Li, J. J. Jasieniak, Y. Li, H. Wu, Y. Luo, D. Li and Q. Meng, Identification of high-temperature exciton states and their phase-dependent trapping behaviour in lead halide perovskites, *Energy Environ. Sci.*, 2018, **11**, 1460-1469.

75  Y. Dong, W. Li, X. Zhang, Q. Xu, Q. Liu, C. Li and Z. Bo, Highly efficient planar perovskite solar cells via interfacial modification with fullerene derivatives, *Small*, 2016, **12**, 1098-1104.

76  B. Duan, L. Guo, Q. Yu, J. Shi, H. Wu, Y. Luo, D. Li, S. Wu, Z. Zheng and Q. Meng, Highly efficient solution-processed CZTSSe solar cells based on a convenient sodium-incorporated post-treatment method, *J. Energy Chem.*, 2020, **40**, 196-203.